\documentclass{article}
\usepackage{subeqn}
\usepackage{graphicx}
\usepackage{natbib}
\usepackage{amssymb}
\bibpunct{(}{)}{;}{a}{}{,}
\let\osection\section
\renewcommand{\section}{\setcounter{equation}{0}\osection}
\renewcommand{\theequation}{\thesection.\arabic{equation}}

\renewcommand{\deg}{\ensuremath{^\circ}} 
\newcommand{\beq}{\begin{equation}}
\newcommand{\eeq}{\end{equation}}
\newcommand{\beqa}{\begin{eqnarray}}
\newcommand{\eeqa}{\end{eqnarray}}
\newcommand{\beqas}{\begin{eqnarray*}}
\newcommand{\eeqas}{\end{eqnarray*}}
\newcommand{\ba}{\left(\begin{array}} 
\newcommand{\ea}{\end{array}\right)}
\newcommand{\bfig}{\begin{figure}}
\newcommand{\efig}{\end{figure}}
\newcommand{\incgcw}[1]{\centering{\includegraphics[width=0.8\columnwidth]{#1}}}
\newcommand{\eps}{\varepsilon}
\newcommand{\av}[1]{\left\langle#1\right\rangle}
\newcommand{\sod}[1]{\,{\rm d}#1} 
\newcommand{\fod}[2]{\frac{{\rm d} #1}{{\rm d} #2}}
\newcommand{\fodn}[3]{\frac{{\rm d}^{#3} #1}{{\rm d} #2^{#3}}}
\newcommand{\dsfrac}[2]{\frac{\displaystyle{#1}}{\displaystyle{#2}}} 
\newcommand{\pfrac}[3]{\left(\frac{#1}{#2}\right)^{#3}}
\newcommand{\rs}[1]{_{\rm #1}} 
\providecommand{\operatorname}[1]{\mathop{\mathrm{#1}}\nolimits}
\renewcommand{\Im}{\operatorname{Im}}
\newcommand{\Fp}{\operatorname{Fp}} 
\newcommand{\am}{\operatorname{am}}
\newcommand{\sn}{\operatorname{sn}}
\newcommand{\cn}{\operatorname{cn}}
\newcommand{\dn}{\operatorname{dn}}
\newcommand{\sd}{\operatorname{sd}}
\newcommand{\e}{{\rm e}} 
\newcommand{\ii}{{\rm i}} 
\renewcommand{\ll}{\!<\!\!<\!} 
\ifx\leqslant\undefined
\else

  \renewcommand{\ge}{\geqslant}
\fi
\newcommand\Ipe{I$\rs{pe}$}
\newcommand\IIpe{II$\rs{pe}$}

\title{Weakly nonlinear waves in magnetized plasma with a slightly
  non-Maxwellian electron distribution. Part 2. Stability of cnoidal~waves}
\author{S. PHIBANCHON$^1$, M. A. ALLEN$^{1,}$\footnote{corresponding
  author} \mbox{} and  G. ROWLANDS$^2$\\
\small $^1$Physics Department, Mahidol University, Rama 6 Road,
Bangkok 10400 Thailand\\
\small $^2$Department of Physics, University of Warwick, Coventry, 
CV4 7AL, UK\\
\small (frmaa@mahidol.ac.th)}
\date{}
\begin{document}
\maketitle
\begin{abstract}
We determine the growth rate of linear instabilities resulting from
long-wavelength transverse perturbations applied to periodic nonlinear
wave solutions to the Schamel-Korteweg-de Vries-Zakharov-Kuznetsov
(SKdVZK) equation which governs weakly nonlinear waves in a strongly
magnetized cold-ion plasma whose electron distribution is given by two
Maxwellians at slightly different temperatures.  To obtain the growth
rate it is necessary to evaluate non-trivial integrals whose number is
kept to minimum by using recursion relations. It is shown that a key
instance of one such relation cannot be used for classes of solution
whose minimum value is zero, and an additional integral must be
evaluated explicitly instead. The \mbox{SKdVZK} equation contains two
nonlinear terms whose ratio $b$ increases as the electron distribution
becomes increasingly flat-topped.  As $b$ and hence the deviation from
electron isothermality increases, it is found that for cnoidal wave
solutions that travel faster than long-wavelength linear waves, there
is a more pronounced variation of the growth rate with the angle
$\theta$ at which the perturbation is applied. Solutions whose minimum
value is zero and travel slower than long-wavelength linear waves are
found, at first order, to be stable to perpendicular perturbations and
have a relatively narrow range of $\theta$ for which the first-order
growth rate is not zero.
\end{abstract}
\section{Introduction}
In Part~1 \citep{APR07} we considered solitary wave solutions of a
modified version of the Zakharov-Kuznetsov (ZK) equation which, in a
frame moving at speed $V$ above the speed of long-wavelength linear waves,
takes the form 
\beq
u_t+(u+bu^{1/2}-V)u_x+\nabla^2u_x=0 \label{e:skdvzk}
\eeq
where the subscripts denote derivatives.  We referred to this equation
as the Schamel-Korteweg-de Vries-Zakharov-Kuznetsov (SKdVZK) equation
as it contains both the quadratic nonlinearity of the KdV equation and
the half-order nonlinearity of the Schamel equation. The equation
governs weakly nonlinear ion-acoustic waves in a plasma permeated by a
strong uniform magnetic field in the $x$-direction. The plasma
contains cold ions and two populations of hot electrons, one free and
the other trapped by the wave potential, whose effective temperatures
differ slightly.  In (\ref{e:skdvzk}) $u$ is proportional to the
electrostatic potential, and $b=(1-T\rs{ef}/T\rs{et})/\sqrt\pi$ where
$T\rs{ef}$ and $T\rs{et}$ are the effective temperatures of the free
and trapped electrons, respectively. As $b$ increases, the electron
distribution 
becomes less peaked. A flat-topped distribution is in accordance
with numerical simulations and experimental observations of
collisionless plasmas \citep{Sch73}.
 For further background to the
physical basis and applicability of the SKdVZK and related equations,
reference should be made to Part~1.

The existence of planar solitary wave solutions to the SKdVZK equation
and their stability to transverse perturbations was addressed in
Part~1.  In this paper we turn to the study of planar cnoidal wave
solutions to the equation. In Sec.~2 we show that a number of families
of cnoidal wave solutions to the one-dimensional form of
(\ref{e:skdvzk}) exist, but not all can be expressed in closed
form. Linear stability analysis of periodic solutions of the SKdVZK
equation with respect to transverse perturbations is carried out in
Sec.~3. Such an analysis has been carried out on cnoidal wave
solutions of the ZK and SZK equations which contain single quadratic and
half-order nonlinearities, respectively \citep{Inf85,MP99}. However, 
as far as we are aware, such a calculation has not been performed
before on an equation containing two nonlinear terms.  The stability
analysis leads to a nonlinear dispersion relation in the form of a
cubic equation whose coefficients are finite-part integrals involving the
unperturbed solution and its derivative. As the solutions contain
elliptic functions the integrals are non-trivial. Recursion relations
between the integrals are derived in order that only the simplest
finite integrals need to be evaluated directly. For some types of
solution it is shown that one instance of a recursion relation cannot be
used and an extra integral must be found directly.
In Sec.~4 we examine how the first-order coefficient of the 
growth rate found from the nonlinear
dispersion relation depends on the type of cnoidal wave, the angle at
which the perturbation is applied, and $b$. Our conclusions are
presented in the final section.

\section{Cnoidal wave solutions}

\begin{figure}[!ht]
\incgcw{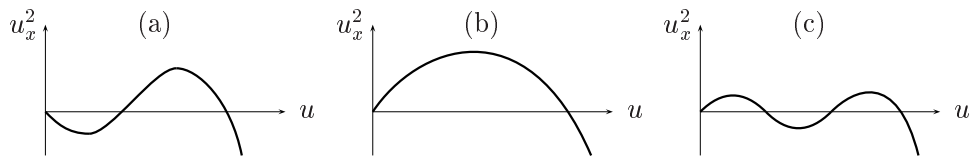}
\caption{\label{f:skdvpp1} $(u,u_x^2)$-sketches of (\ref{e:ux2}) with
  $C_0=0$ showing existence of families of periodic wave solutions:
(a)~$C<0$, $V>0$ or $b<0$ (or both); (b)~$C>0$; (c)~$C>0$ for some
  $V<0$ and $b<0$.}
\end{figure}

To look for planar cnoidal wave solutions of permanent form travelling
at speed $V$ above the long-wavelength linear wave speed we drop the
$t$, $y$ and $z$ dependence in (\ref{e:skdvzk}).
Integrating once then gives
\beq
u_{xx}=\frac{C}{2}+Vu-\frac{2}{3}bu^{3/2}-\frac{1}{2}u^2, \label{e:uxx}
\eeq
and multiplying by $2u_x$ and integrating once more yields
\beq
u_x^2=C_0+Cu+Vu^2-\frac{8}{15}bu^{5/2}-\frac{1}{3}u^3 \label{e:ux2}
\eeq
where $C_0$ and $C$ are integration constants. 
Although from phase plane analysis it is clear
that a number of families of periodic nonlinear waves exist, only when
$C_0=0$ can closed-form solutions be obtained in general. Sketches of
(\ref{e:ux2}) for the various cases leading to periodic solutions when
$C_0=0$ are shown in Fig.~\ref{f:skdvpp1}.
 
\begin{figure}[!ht]
\incgcw{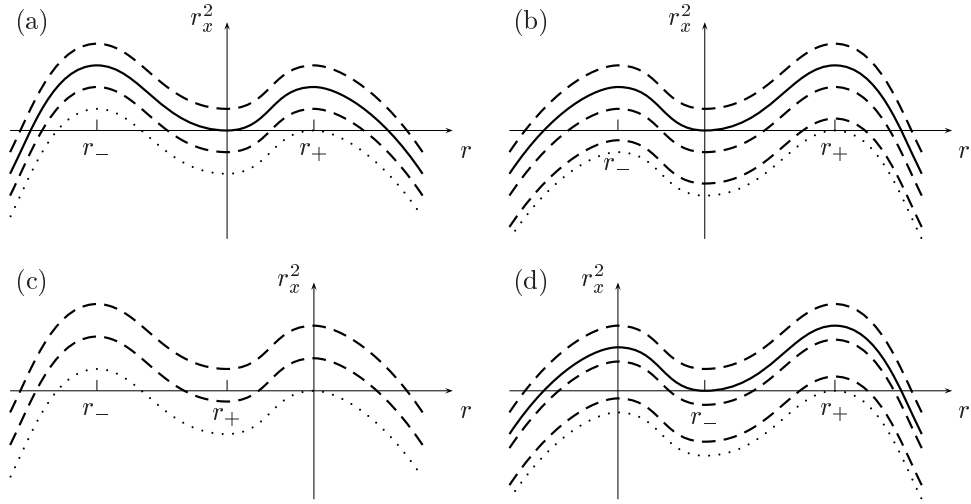}
\caption{\label{f:skdvpp2} $(r,r_x^2)$-sketches of (\ref{e:rx2}) 
  for (a)~$V>0$, $b>0$
  (b)~$V>0$, $b<0$ (c)~$-6b^2/25<V<0$, $b>0$ (d)~$-6b^2/25<V<0$,
  $b<0$. Solid, dashed and dotted curves give rise to solitary wave, periodic
  nonlinear wave, and constant (linear limit) solutions, respectively.}
\end{figure}

After introducing the variable $r=\sqrt{u}$, (\ref{e:ux2}) with
$C_0\equiv0$ reduces to
\beq
4r_{x}^2=g(r)\equiv h(r)+C, 
\qquad h(r)\equiv Vr^2-\frac{8b}{15}r^3-\frac{1}{3}r^4.
\label{e:rx2}
\eeq
Possible forms of $g(r)$ for various $V$, $b$ and $C$ are sketched in
Fig.~\ref{f:skdvpp2}.  The $u^{1/2}$ term that appears in (\ref{e:skdvzk})
must be interpreted as the positive square root and as a result we
must restrict solutions of (\ref{e:rx2}) to $r\ge0$.  In view of this,
at first sight it would appear that nonlinear wave solutions to
(\ref{e:rx2}) that cross the line $r=0$ with positive $r_x^2$ would
have to be discarded. However, since $u_x^2=r^2g(r)$, the $r\ge0$ part
of such a solution forms a complete closed loop that touches the
origin in the ($u$$\ge$$0,u_x)$-plane, as can be seen to occur in
Fig.~\ref{f:skdvpp1}(b) and (c), and hence corresponds to a
nonlinear wave solution with a minimum value of zero.
Although for such solutions in the $(r,r_x)$-plane $r_x$ jumps from a
negative to an equal and opposite positive value at $r=0$, it is
easily shown that $u$ and its derivatives are continuous there. The
jump in the solution in the $(r,r_x)$-plane means that the solutions
are most simply expressed as functions extended by periodicity.
Such solutions have already categorized for the Schamel equation
(which contains the single half-order nonlinearity) in
\citet{OP97}. \citet{Sch72}
described them for the current equation but in the
following they are presented in a more unified form. 

The quartic $g(r)$ will always
have a stationary point at $r=0$ and also at
\beq
r_\pm=\pm\sqrt{\left(\frac{3b}{5}\right)^2+\frac{3V}{2}}-\frac{3b}{5}
\label{e:rpm}
\eeq
provided that $V>-6b^2/25$. From the sketches of $g(r)$ in
Fig.~\ref{f:skdvpp2}, it is apparent that if $r_+$ is real and positive,
nonlinear wave solutions will only occur if $C>-h(r_+)$ and the linear
wave limit corresponds to $C=-h(r_+)$. 

If $g(r)$ has four real roots, $r_1<r_2<r_3<r_4$, 
then solving (\ref{e:rx2}) yields
the cnoidal wave solution
\beq
u(x)= [r(x)]^2 = \left(
\frac{r_4+r_1\rho\sn^2(\eta(x-x_0)|m)}
{1+\rho\sn^2(\eta(x-x_0)|m)} \right)^2   
\label{e:I}
\eeq
where 
\[
\rho\equiv\frac{r_4-r_3}{r_3-r_1}, \qquad
m\equiv\frac{r_2-r_1}{r_4-r_2}\,\rho, \qquad 
\eta \equiv\sqrt{\frac{(r_4-r_2)(r_3-r_1)}{48}},
\]
and $x_0$ is an arbitrary phase. We will refer to this class of
solution as being of type~I. 
Note that when $V>0$, in the soliton limit
$c=0$ we have $r_3=r_2=0$ and (\ref{e:I}) then
reduces to the conventional solitary wave solution given in Part~1.

When $b<0$, it can be seen from Fig.~\ref{f:skdvpp2}(b) and (d) that there
are periodic wave solutions corresponding to a $g(r)$ with just two
real roots for $-h(r_+)<C<-h(r_-)$ when $V>0$, and for $-h(r_+)<C<0$ when
$-6b^2/25<V<0$. If the two real roots are $r\rs{u}$ and $r\rs{l}$,
with $r\rs{u}>r\rs{l}$, and the remaining
two complex conjugate roots are $\alpha\pm\ii\beta$, then solving
(\ref{e:rx2}) in this case gives what we call the type~II solution,
\beq
u(x)= [r(x)]^2=\left( 
\dsfrac{(Ar\rs{l}+Br\rs{u})-(Ar\rs{l}-Br\rs{u})\cn(\bar\eta(x-x_0)|\bar{m})}
{(A+B)-(A-B)\cn(\bar\eta(x-x_0)|\bar{m})}\right)^2,
\label{e:II}
\eeq
where
\[
A =\sqrt{(r\rs{u}-\alpha)^2+\beta^2}, \qquad 
B =\sqrt{(r\rs{l}-\alpha)^2+\beta^2},
\]
and 
\[
\bar{m}=\frac{(r\rs{u}-r\rs{l})^2-(A-B)^2}{4AB}, \qquad
\bar\eta = \sqrt{\frac{AB}{12}}.
\]

We now turn our attention to the solutions written as periodically
extended functions. For $V>0$ these
occur when $c>0$. As in these cases there are only two real roots,
these solutions take a similar form to (\ref{e:II}). However, due to the jump
in the $(r,r_x)$-plane they must be written in the form
\beq
u(x)=\left(
\dsfrac{(Ar\rs{l}+Br\rs{u})-(Ar\rs{l}-Br\rs{u})
\cn(\bar\eta\check{x}(\bar\chi/\bar\eta)|\bar{m})}
{(A+B)-(A-B)\cn(\bar\eta\check{x}(\bar\chi/\bar\eta)|\bar{m})}\right)^2,
\quad
\bar\chi=\cn^{-1}\left(\frac{Ar\rs{l}+Br\rs{u}}{Ar\rs{l}-Br\rs{u}}\right)
\label{e:IIpe}
\eeq
where
\[
\check{x}(p)\equiv(x-x_0+p \bmod 2p) - p.
\]
These solutions have a period of $2\bar\chi/\bar\eta$ and for a given
value of $V$ and $b$ have a larger amplitude than the solitary
wave. We call these type~\IIpe\ solutions. When $b=0$ they reduce
to an ordinary KdV equation cnoidal wave solution with a minimum value of zero.

From Fig.~\ref{f:skdvpp2}(c) and (d) 
it is clear that when $V<0$, 
smaller amplitude solutions that touch $u=0$
are also possible. When there are two real roots,
(\ref{e:IIpe}) still applies, while if $b>0$ and there are four
real roots, what we will refer to as the type~\Ipe\
solution results. It is similar to (\ref{e:I}) but must be written as
\beq
u(x) = \left(
\frac{r_4+r_1\rho\sn^2(\eta\check{x}(\chi/\eta)|m)}
{1+\rho\sn^2(\eta\check{x}(\chi/\eta)|m)} \right)^2,   
\quad
\chi=\sn^{-1}\sqrt{\frac{r_4}{r_1\rho}}.
\label{e:Ipe}
\eeq
When there are four real roots and $b<0$, the solution which has a
minimum value of zero is
the same as the above after making the interchanges $r_4\leftrightarrow
r_2$ and $r_3\leftrightarrow r_1$.

\section{Linear stability analysis}

By using the small-$k$ expansion
method \citep{Row69,Inf85,IR-NWSC},
we now investigate the linear stability of periodic waves to 
long-wavelength perturbations with wavevector
$k(\cos\theta,\sin\theta\cos\varphi,\sin\theta\sin\varphi)$ 
where $\theta$ is the angle between the
direction of the wavevector and the $x$-axis, and $\varphi$ is the
azimuthal angle. We start from the ansatz
\beq
u=u_0(x)+\eps\Phi(x)\e^{\ii
  k(x\cos\theta+y\sin\theta\cos\varphi+z\sin\theta\sin\varphi)-\ii\omega t}
\label{e:uperturb}
\eeq
where $u_0(x)$ is a periodic solution to (\ref{e:skdvzk}), $\eps\ll1$,
and the eigenfunction $\Phi(x)$ must have the same period as
$u_0(x)$. 
Substituting (\ref{e:uperturb}) into
(\ref{e:skdvzk}) and linearizing with respect to $\eps$ gives
\beq
\fod{}{x}L\Phi=\ii\omega\Phi-\ii k\cos\theta\,Q\Phi
-3\ii k\cos\theta\,\Phi_{xx}+k^2(1+2\cos^2\theta)\Phi_x+\ii k^3\cos\theta\,\Phi
\label{e:LPhix}
\eeq
where 
\[
L\equiv\fodn{}{x}{2}+Q, \qquad Q\equiv u_0+bu^{1/2}_0-V,
\]
and $\Phi$ and $\omega$ are written as the expansions,
\beqa
\Phi &=& \Phi_0+k\Phi_1+ \ldots, \label{e:Phi}\\
\omega&=&\omega_1k+\omega_2k^2+\ldots\;.
\label{e:ome}
\eeqa
For the remainder of the calculation we follow a similar procedure to that
first given in \citet{Par93}.
After substituting (\ref{e:Phi}) and (\ref{e:ome}) into
(\ref{e:LPhix}) and equating coefficients of $k^n$ we obtain
the sequence of equations
\beq
(L\Phi_n)_x = R_{nx}(x) \label{e:LPhin}
\eeq
in which the expressions for $R_{nx}$ are of the same form as
in Part~1 after replacing $\gamma_j$ by $-\ii\omega_j$. 
Since $Lu_{0x}=0$, the solution to $L\Phi_n=R_n+B_n$, where $B_n$
are integration constants obtained on integrating (\ref{e:LPhin}), is
\beq
\Phi_n=u_{0x}v_n \label{e:Phin}
\eeq
where
\beq
v_{nx}=\frac{1}{u_{0x}^2}
\left\{A_n+\int^x\!\!\!(R_n(x')+B_n)u_{0x}(x')\sod{x'}\right\}
\label{e:vnx}
\eeq
and $A_n$ are additional constants. On integrating (\ref{e:vnx}), secular
(non-periodic) terms will occur in $v_n$. To remove these we
must insist that 
\beq
\av{v_{nx}}=0 \label{e:avvnx}
\eeq
where 
\beq
\av{f}=\Fp\frac{1}{\lambda}\int_0^\lambda \!\!\!f(x)\sod{x},
\eeq
$\lambda$ is the period of $u_0$, and $\Fp$ stands for
Hadamard's finite part \citep{Zem-DTTA}. Equation (\ref{e:avvnx})
provides a relation between $A_n$ and $B_n$ which can later be used to help
eliminate these constants. 

To lowest order in $k$ we have
$(L\Phi_0)_x=0$. As a result of the translational invariance of $u_0$,
this has a solution proportional to $u_{0x}$. This result can be
obtained more explicitly, as is done in \citet{MP99}, by using the
consistency conditions $\av{v_{0x}}=0$, $\av{(L\Phi_1)_x}=0$, and 
$\av{u_0(L\Phi_1)_x}=0$ to show that $v_{0x}=0$. Without loss of
generality we choose a unit
constant of proportionality (which corresponds to setting $v_0=1$) and
we are left with
\beq
\Phi_0=u_{0x}. \label{e:Phi0}
\eeq

Integrating the first-order equation gives
\beq
L\Phi_1=\ii\omega_1u_0-2\ii\cos\theta\,u_{0xx}+ B_1, \label{e:LPhi1}
\eeq
and after using (\ref{e:vnx}) one obtains 
\beq
v_{1x}=\frac{1}{u_{0x}^2}
\left(A_1+\frac{\ii\omega_1u_0^2}{2}-\ii\cos\theta\ u_{0x}^2+ B_1u_0\right).
\label{e:v1x}
\eeq
Then applying (\ref{e:avvnx}) results in the relation
\beq
A_1\beta_0+\frac{\ii\omega_1\beta_2}{2}-\ii\cos\theta+ B_1\beta_1=0 
\label{e:avv1x}
\eeq
in which we have introduced the quantities
\beq
\beta_s \equiv \av{\frac{u_0^s}{u_{0x}^2}}. \label{betan}
\eeq

After using (\ref{e:Phin}) and (\ref{e:LPhi1}), the second-order
equation may be written as  
\beq
(L\Phi_2)_x=
\ii\omega_2u_{0x}+\ii\omega_1u_{0x}v_1+u_{0xx}+\omega_1\cos\theta\,u_0
-\ii B_1\cos \theta -2\ii\cos\theta\, (u_{0x}v_1)_{xx}.
\label{e:LPhi2} 
\eeq
To obtain $\omega_1$ it is not necessary to evaluate $\Phi_2$. Instead,
we first apply the finite-part averaging operation $\av{\cdot}$ to
(\ref{e:LPhi2}). After using partial integration to show that 
 $\av{u_{0x}v_1}=-\av{u_0v_{1x}}$, and by virtue of the periodicity of
$\Phi_2$ (which implies that $\av{(L\Phi_2)_x}=0$), we get
\beq
-\ii\omega_1\av{u_0v_{1x}}+\omega_1\cos\theta\,\alpha_1-\ii B_1\cos\theta=0 
\label{e:avf2}
\eeq
where we have defined
\beq
\alpha_s \equiv \av{u_0^s}.
\label{e:avalp}
\eeq
We then multiply (\ref{e:LPhi2}) by $u_0$ and apply $\av{\cdot}$. The
left-hand side can be shown to be zero by integrating by parts and
then using the self-adjoint property of $L$ and the fact that
$Lu_{0x}=0$. This leaves, after further manipulation via partial integration,
\beq
-\frac{\ii\omega_1}{2}\av{u_0^2v_{1x}}-\av{u_{0x}^2}
+\omega_1\cos\theta\,\alpha_2
-\ii B_1\cos\theta\,\alpha_1
+\ii\cos \theta \av{u_{0x}^2 v_{1x}} =0. 
\label{e:avu0f2}
\eeq
From (\ref{e:v1x}) we can obtain
\begin{subeqnarray}\label{e:avuv1x}
\av{u_0v_{1x}}&=&A_1\beta_1+\frac{\ii\omega_1\beta_3}{2}
-\ii\cos\theta\,\alpha_1+B_1\beta_2,\\
\av{u_0^2v_{1x}}&=&A_1\beta_2+\frac{\ii\omega_1\beta_4}{2}
-\ii\cos\theta\,\alpha_2+B_1\beta_3,\\
\av{u_{0x}^2v_{1x}}&=&A_1+\frac{\ii\omega_1\alpha_2}{2}
-\ii\cos\theta\av{u_{0x}^2}+B_1\alpha_1,
\end{subeqnarray}
and after replacing $u$ in (\ref{e:ux2}) by $u_0$ and applying
$\av{\cdot}$ we have
\[
\av{u_{0x}^2}
=C\alpha_1+V\alpha_2-\frac{8b}{15}\alpha_{5/2}-\frac{1}{3}\alpha_3.
\]
Substituting (\ref{e:avuv1x}) into (\ref{e:avf2}) and (\ref{e:avu0f2})
and then eliminating $A_1$ and $B_1$ from these two equations and 
(\ref{e:avv1x}) leaves the following equation for $\omega_1$
\beq
a_0+a_1\omega_1+a_2\omega_1^2+a_3\omega_1^3 = 0, \label{e:nondisp}
\eeq
where
\beqas
  a_0 &=& (\beta_0\av{u_{0x}^2}\sin^2\theta+\cos^2\theta)\cos\theta,\\ 
  a_1 &=& (\beta_0\beta_2-\beta_1^2)\av{u_{0x}^2}\sin^2\theta ,\\
 a_2 &=& \left(\beta_1\beta_3-\frac34\beta_2^2
-\frac14\beta_0\beta_4\right)\cos\theta ,\\
 a_3 &=& \frac{1}{4}(\beta_1^2\beta_4+\beta_2^3+\beta_0\beta_3^2
-\beta_0\beta_2\beta_4-2\beta_1 \beta_2 \beta_3).
\eeqas

Owing to the fact that $u_{0x}$ is zero at some points, the direct
evaluation of the $\beta_s$ would require a finite-part
calculation. This can be avoided by instead expressing these
quantities in terms of the $\alpha_s$. To accomplish this, we require
a number of recursion relations. The first of these is obtained by
multiplying (\ref{e:uxx}) by $u_0^s/u_{0x}^2$, applying $\av{\cdot}$,
and then simplifying the left-hand side using partial
integration. This yields
\beq
s\alpha_{s-1}=\frac{C}{2}\beta_s+V\beta_{s+1}-\frac{2b}{3}\beta_{s+3/2}
-\frac12\beta_{s+2}. 
\label{e:beta1}
\eeq
Applying the same procedure to (\ref{e:ux2}) and then replacing $s$ by
$s-1$ gives
\beq
\alpha_{s-1}=C\beta_s+V\beta_{s+1}-\frac{8b}{15}\beta_{s+3/2}
-\frac13\beta_{s+2}.
\label{e:beta2}
\eeq
Eliminating $\beta_{s+3/2}$ from the above two equations gives
\beq
9C\beta_s+3V\beta_{s+1}+\beta_{s+2}=3(5-4s)\alpha_{s-1}, \label{e:bl1}
\eeq  
and putting the values $s=0,1,2$ into (\ref{e:bl1}) generates the
following three equations involving the required $\beta_s$:
\begin{subeqnarray}\label{e:3b}
9C\beta_0+3V\beta_1+\beta_2 &=& -15\alpha_{-1}, \label{e:b012}\\   
9C\beta_1+3V\beta_2+\beta_3 &=& 3, \label{e:b123}\\   
9C\beta_2+3V\beta_3+\beta_4 &=& -\alpha_1. \label{e:b234}
\end{subeqnarray}
A further two equations for the $\beta_s$ are found by first
eliminating $\beta_{s+2}$ from (\ref{e:beta1}) and (\ref{e:beta2}) 
to give
\beq
30C\beta_s+15V\beta_{s+1}-4b\beta_{s+3/2}=15(3-2s)\alpha_{s-1}.
\label{e:bl2}
\eeq
Putting $s=1$ and $s=3/2$ into this equation and eliminating
$\beta_{5/2}$, and then using the resulting equation and (\ref{e:bl2})
with $s=0$ to eliminate $\beta_{3/2}$ gives
\beq
15C^2\beta_0+15VC\beta_1+\frac{15V^2\beta_2}{4}-\frac{4b^2\beta_3}{15} = 
\frac{45}{2}C\alpha_{-1}+\frac{15V}{4}.
\label{e:b0123} 
\eeq
Eliminating $\beta_{7/2}$ and $\beta_{5/2}$ from the equations
obtained from (\ref{e:bl2}) with $s=1,2,5/2$ yields
\beq
15C^2\beta_1+15VC\beta_2+\frac{15V^2\beta_3}{4}-\frac{4b^2\beta_4}{15} = 
\frac{15}{2}C-2b\,\alpha_{3/2}-\frac{15V\alpha_1}{4}. \label{e:b1234}
\eeq
Using (\ref{e:3b}), (\ref{e:b0123}) and (\ref{e:b1234}), all the
$\beta_s$ for $s=0,\ldots,4$ can then be expressed in terms of $\alpha_1$,
$\alpha_{-1}$ and $\alpha_{3/2}$.

We now turn to the evaluation of $\alpha_s$. A recursion relation
involving only $\alpha_s$ can be obtained by 
multiplying (\ref{e:uxx}) by $u_0^s$ and (\ref{e:ux2}) by
$su_0^{s-1}$, adding, and then averaging. If $u_0>0$ or if $s\ge0$,
the average value of $u_0^s$ will be finite and equal to $\alpha_s$
and we may then write
\beq
\left(\frac12+\frac{s}{3}\right)\alpha_{s+2}=
\left(\frac12+s\right)C\alpha_s
+(1+s)V\alpha_{s+1}-\left(\frac23+\frac{8s}{15}\right)b\,\alpha_{s+3/2}.
\label{e:alpha}
\eeq
However, if $u_0(x)$ is zero at some values of $x$, as is the case for the
type~\Ipe\ and \IIpe\ solutions, and $s<0$, the average of
$u_0^s$ will no longer be finite, and in cases where the coefficient of
an infinite integral is zero, (\ref{e:alpha}) has to be modified.
Before continuing, 
it should be noted that, in contrast, (\ref{e:beta1}) is always valid
for $s=0$ since for this value of $s$ the left-hand side originates
from $\av{u_{0xx}/u_{0x}^2}$ which is identically zero.

From (\ref{e:alpha}) it is evident that we will have to evaluate at
least two of the $\alpha_s$ directly. The simplest to find, due to the
fact that the periodic wave solutions are of the form
$u_0(x)=[r(x)]^2$, are $\alpha_{1/2}$ and $\alpha_1$. The evaluation
of these integrals for type~I, \IIpe, and \Ipe\ solutions is
given in the Appendix.  To determine the $\beta_s$ and $\av{u_{0x}^2}$
we also require $\alpha_{-1}$, $\alpha_{3/2}$, $\alpha_2$,
$\alpha_{5/2}$, and $\alpha_3$. Putting $s=-1$ in (\ref{e:alpha})
presents no problem as the only coefficient that is zero is
multiplying a term originating from a finite integral. Thus we have
\[
\alpha_{-1}=-\frac1{3C}\left(\alpha_1+\frac{4b\alpha_{1/2}}{15}\right).
\]
To find $\alpha_{3/2}$ we need to use $s=-1/2$. In this case
(\ref{e:alpha}) must be re-written in the form
\beq
\alpha_{3/2} = \frac{3V\alpha_{1/2}}{2}-\frac{6b\alpha_1}{5}
+\lim_{s\to-1/2}\Fp\frac{3(1+2s)}{2\lambda}\int_0^\lambda \!\!\!u_0^s\sod{x}.
\label{e:a32}
\eeq
For type~I solutions, the final term on the right-hand side 
 is zero. For the type~\Ipe\ and \IIpe\
solutions, the integral in (\ref{e:a32}) is infinite and the finite
part would have to be found numerically. In such cases $\alpha_{3/2}$
needs to be obtained directly, as is done in the Appendix. The remaining
$\alpha_s$ can be found in a straightforward manner by putting
$s=0$, $\frac12$, and 1 into (\ref{e:alpha}) which give, respectively,
\beqas
\alpha_2&=&C+2V\alpha_1-\frac{4b\alpha_{3/2}}{3}, \\
\alpha_{5/2}&=&\frac{3C\alpha_{1/2}}{2}+\frac{9V\alpha_{3/2}}{4}
-\frac{21b\alpha_2}{15},\\ 
\alpha_3&=&
\frac{9C}{5}\alpha_1+\frac{12V}{5}\alpha_{2}-\frac{36b\alpha_{5/2}}{25}.
\eeqas
The values of $\alpha_s$ corresponding to finite integrals 
 obtained using the procedure
outlined were checked by numerical integration for specific values of
the parameters. The numerical values of the 
remaining quantities, namely, $\alpha_{-1}$ for the periodically
extended solutions and the $\beta_s$,
 for which the finite-part
operation is not redundant, were checked using a finite-part
numerical integration technique \citep{OKe-93,Phi-06}. 

\section{Growth rate of instabilities}

Having obtained the three roots to the nonlinear dispersion relation
(\ref{e:nondisp}), we discard the real parts as they are of no
importance in the context of stability. 
The solution is unstable if two of the roots are
 complex conjugates. If $\omega_1$ is one of these roots,
the first-order growth rate of the instability is given by
\[
\gamma\equiv\gamma_1k\equiv|\Im\omega_1|\,k.
\]

When examining the dependence of the growth rate on the type of
solution and the direction of the perturbation we find it 
convenient
to introduce the parameter $c$, a rescaled version of $C$, defined by
\beq
c\equiv\frac{C}{|h(r_+)|},
\eeq
provided that $V>-6b^2/25$. 
Then, if $V>0$, the linear limit corresponds to $c=-1$ and the soliton
limit occurs at $c=0$. The type \Ipe\ and \IIpe\ solutions
have $c>0$. As in Part~1, we only consider the stability of
solutions for which $b>0$ as these are the more physically relevant.

We feel that a plot of $\gamma_1$ against $\theta$ shows the angular
dependence of the growth rate more clearly than the more traditional
approach of using a polar plot to depict the dependence of the real
and imaginary parts of $\omega$ at all angles.  The value of
$\gamma_1$ for type I solution instabilities as a function of angle
for a number of values of $c$ and $b$ are shown in
Fig.~\ref{f:grthI}. For the soliton limit (when $c=0$) the growth rate
is proportional to $\sin\theta$ which is in agreement with the results
of Part 1. For the cnoidal wave solutions (when $-1<c<0$), $\gamma_1$
is only non-zero above a critical angle, $\theta\rs{crit}$, which
increases with decreasing $c$. It is also evident that
$\theta\rs{max}$, the angle at which the maximum growth rate occurs,
differs from 90$\deg$ for cnoidal waves. The variation of both
$\theta\rs{max}$ and $\theta\rs{crit}$ with $c$ is shown in
Fig.~\ref{f:trI2IIp}(a). The growth rate is largest for the soliton
limit. From the plot of $\gamma\rs{1m}$, the maximum value of
$\gamma_1$ (the value when $\theta=\theta\rs{max}$), in
Fig.~\ref{f:grI2IIp}(a), it is apparent that there is a rapid variation in
growth rate as $c$ approaches zero. This is not unexpected given that
the waveform period increases rapidly and becomes infinite at the
soliton limit, $c=0$. Notice that the results found for the soliton
limit are in agreement with the analytical results
given in Part 1 of this study.

\bfig[!ht]
\incgcw{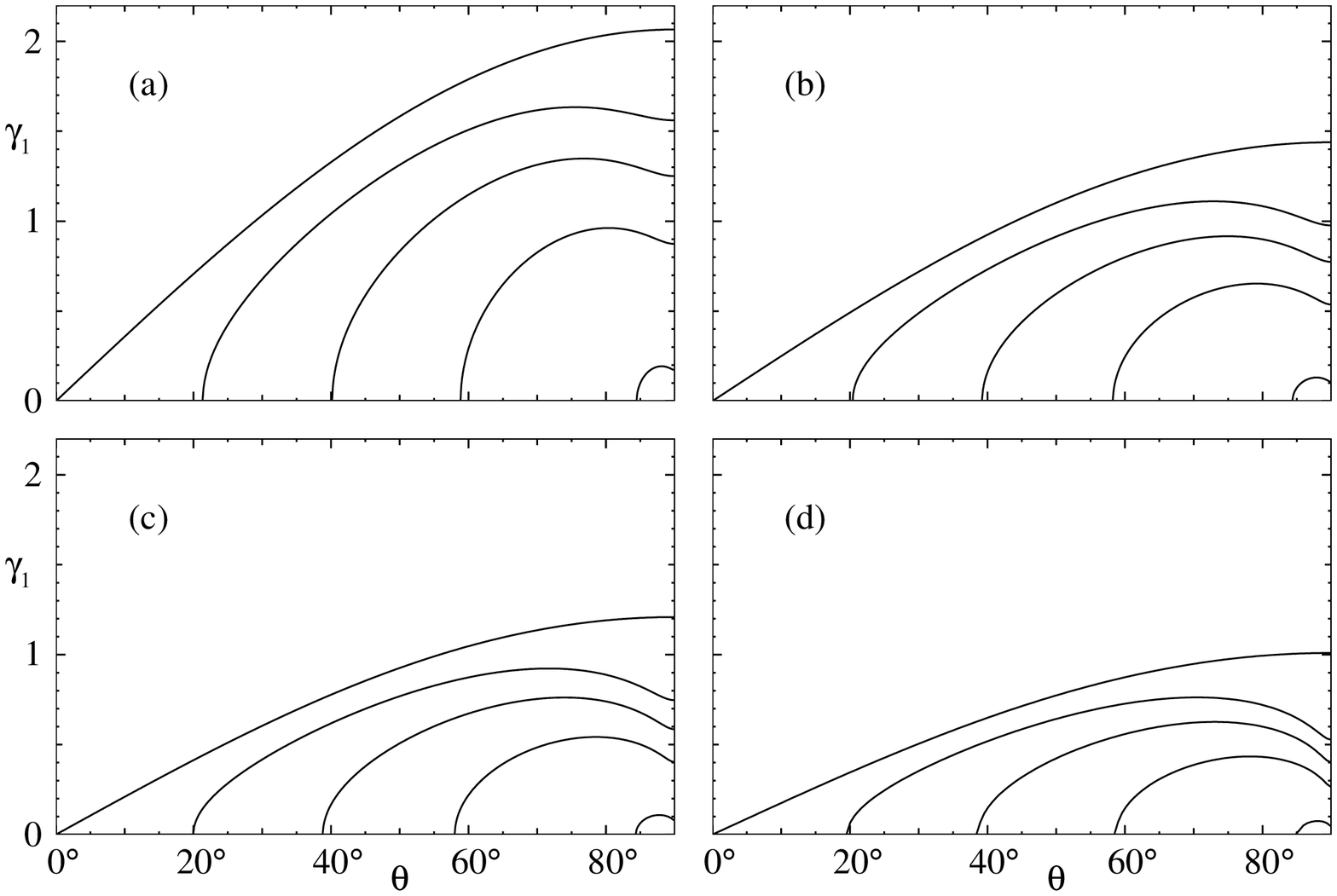}
\caption{Plots of $\gamma_1$ against $\theta$ for type~I solutions
  with $V=4$ and $c$ of $0$ (top curves), $-0.25$,
  $-0.5$, $-0.75$, and $-0.99$ (lowest curves). (a)~$b=0$ (b)~$b=2$
  (c)~$b=4$ (d)~$b=50$.}
\label{f:grthI}
\efig

\bfig[!ht]
\incgcw{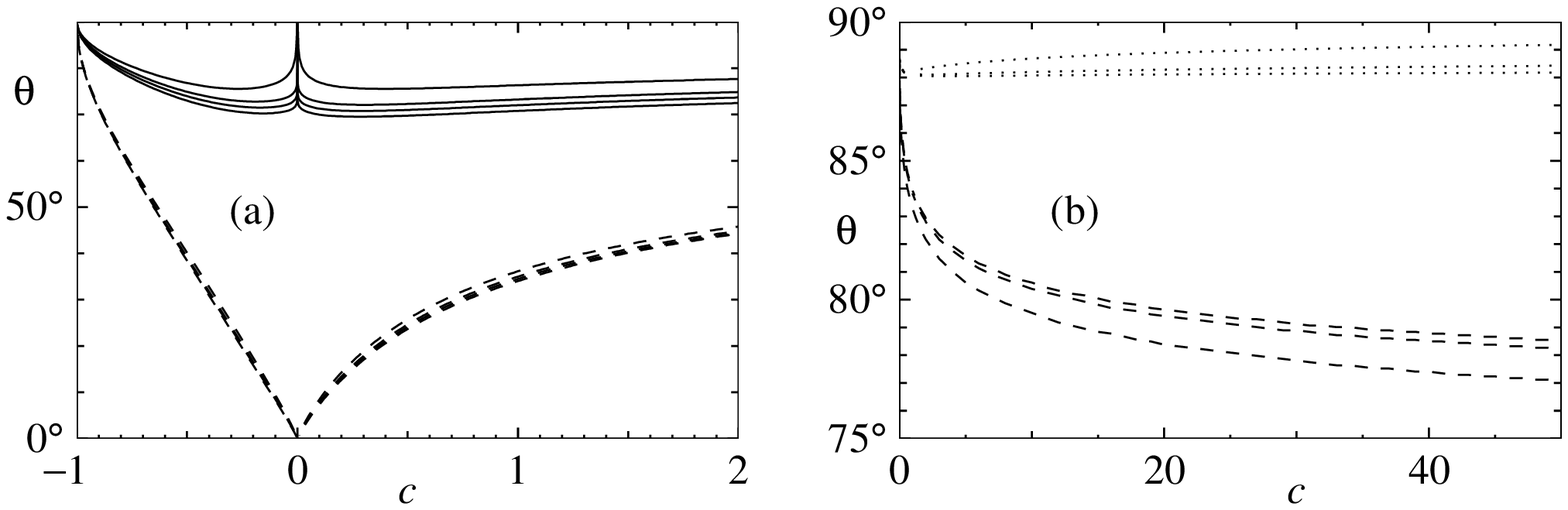}
\caption{Plots of $\theta\rs{max}$ [solid lines], $\theta\rs{crit}$
  [dashed lines], and $\theta\rs{c}$ [dotted lines] against $c$:
  (a)~$V=4$ with $b$ taking the values $0$ (top curves), $2$, $4$, and
  $50$ (lowest curves); (b)~$V=-4$ with $b$ taking the values $5$
  (outermost curves), $10$, and $20$ (innermost curves).}
\label{f:trI2IIp}
\efig

\bfig[!ht]
\incgcw{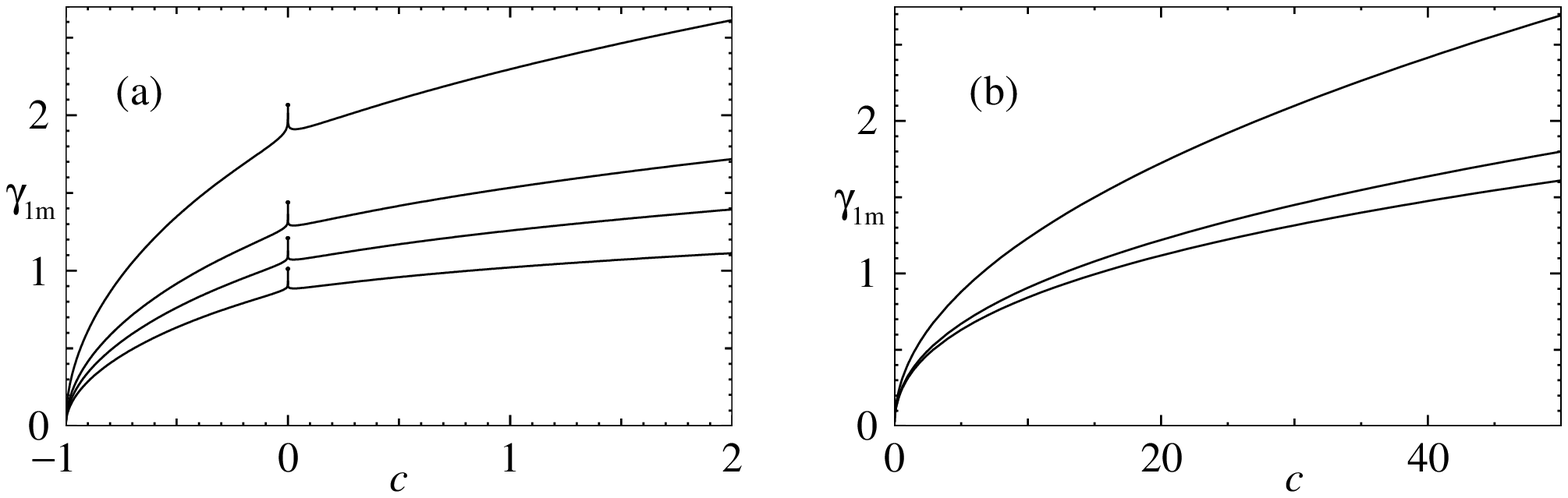}
\caption{Plots of $\gamma\rs{1m}$ against $c$:
  (a)~$V=4$ with $b$ taking the values $0$ (top curves), $2$, $4$, and
  $50$ (lowest curves) -- dots indicate value of $\gamma_1$ calculated
  for solitary wave solutions in Part~1; (b) $V=-4$ with $b$ taking
  the values $5$ (top curves), $10$, and $20$ (lowest curves).}
\label{f:grI2IIp}
\efig

In Part 1 it was found that $\gamma_1$ for solitary waves decreases
with increasing $b$ for a fixed value of $\eta$. As is apparent from
equation (2.2) of Part 1, for fixed $\eta$, the amplitude decreases as
$b$ increases. However, if the amplitude is fixed (by using the
appropriate value of $\eta$ in each case) then it is found that
$\gamma_1$ increases with $b$. Cnoidal waves for different values of $b$
but with the same amplitude and values of $m$ will have different
periods. It therefore seems inappropriate to compare the growth
rates in such cases. Nevertheless, meaningful comparisons can be made
on examining the entire growth rate curve as a function of $\theta$. 
As can be seen from Figs.~\ref{f:grthI} and \ref{f:grI2IIp}(a),
 there is a more marked
variation of the growth rate with $\theta$ for angles above
$\theta\rs{max}$ as $b$ increases, and the plots in
Fig.~\ref{f:trI2IIp}(a) indicate that $\theta\rs{max}$ deviates from the
perpendicular most of all when $b$ is large. On the other hand,
$\theta\rs{crit}$ shows only a slight dependence on $b$.

We now turn to the stability results for solutions in the form of
functions extended by periodicity. When $V>0$ and $c$ is increased
above zero, one obtains type~\IIpe\ solutions. It can be seen from
Fig.~\ref{f:IIpIp}(a)-(c) that the first-order growth rates of these solutions
are higher than that for the soliton limit at some angles, but this
range of angles decreases with increasing $b$. In addition to an
increasing $\theta\rs{crit}$ with $c$, the first-order growth rate for
exactly perpendicular perturbations vanishes for large enough $b$ and $c$.
Evidently the 
growth rate has a significantly greater angular dependence than for
the Type~I solutions.

\bfig[!ht]
\incgcw{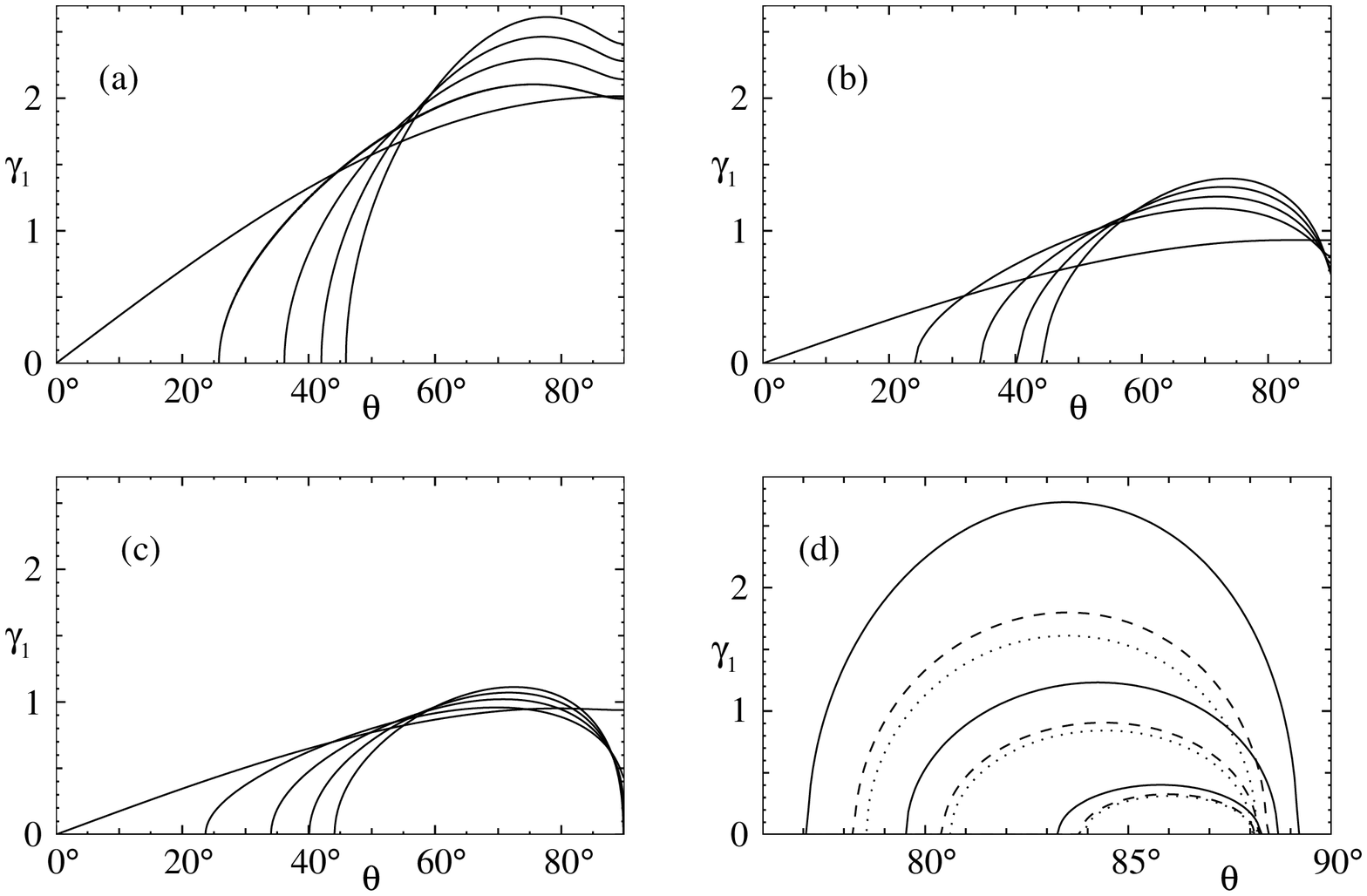}
\caption{Plots of $\gamma_1$ against $\theta$ for type~\IIpe\ and
  \Ipe\ solutions. In (a)-(c), $V=4$ and $c$ takes the values $0$
  (leftmost curves), $0.5$, $1.0$, $1.5$ and $2.0$ (rightmost curves)
  and (a)~$b=0$ (b)~$b=4$ (c)~$b=50$. In (d) $V=-4$ and $c$ takes the
  values $50$ (top curves), $10$, and $1$ (lowest curves) with $b=5$
  [solid lines], $b=10$ [dashed lines], and $b=20$ [dotted lines].}
\label{f:IIpIp}
\efig

For the stability results we have examined so far, the first-order
growth rate is non-zero for angles just below 90$\deg$. In the case of
 type~\Ipe\
and \IIpe\ solutions when $V<0$ the results in Fig.~\ref{f:IIpIp}(d)
indicate that there is a cut-off angle $\theta\rs{c}$ above which
$\gamma_1$ vanishes. Hence such waves are, to first order, stable to
perpendicular perturbations. In addition, the instability occurs over
a relatively small range of angles, even for large values of $c$. 
For these types of solution, as is shown in Fig.~\ref{f:grI2IIp}(b), the
growth rate increases monotonically with $c$, 
in contrast to the behaviour near the
type~I to type~\IIpe\ transition. There is no spike in the growth
rate at $c=1$, the type~\Ipe-\IIpe\ transition, since there
is no sudden change of period around that point.

\section{Conclusions}

This article has dealt with the small-$k$ stability with respect to
transverse perturbations of cnoidal wave solutions to the SKdVZK
equation which governs strongly magnetized plasma with slightly
non-isothermal electrons.  It was found that the growth rate of
instabilities for ordinary cnoidal waves that travel faster than
long-wavelength linear waves has a stronger angular variation as the
distribution of electrons becomes increasingly flatter than the
isothermal Maxwellian.  We also examined a class of solutions that do
not occur for equations without a square root term. These solutions,
which are written as functions extended by periodicity, have a minimum
value of zero. This causes difficulties with some instances of
recursion relations used in the determination of the growth rate, and
results in it being necessary to evaluate an additional integral.
This type of solution, for the case when the wave velocity is less
than that of long-wavelength linear waves, has, to first order, a
relatively narrow range of perturbation angles at which instability
occurs and is stable to perpendicular perturbations.

The half-integer nonlinear term in the SKdVZK equation
 was originally introduced by Schamel
to model the effect of trapped particles in Bernstein-Greene-Kruskal
(BGK) solutions of the Vlasov-Poisson equation. Hence Part 1 and this
paper are to be viewed as a step towards the more formidable problem
of studying the stability of the BGK modes where trapped particles play
a significant role.

\section*{\normalsize \em Acknowledgements}
The authors wish to gratefully acknowledge support from the Thai
Research Fund (PHD/47/2547). Two of the authors (S.P. and M.A.A.)
also thank Warwick University for its hospitality during their visits.

\appendix
\section{The evaluation of $\alpha_{1/2}$, $\alpha_1$ and $\alpha_{3/2}$}
For cnoidal wave solutions, $\alpha_{1/2}$, $\alpha_1$, and
$\alpha_{3/2}$ are elliptic integrals and for their evaluation 
we therefore rely heavily on
\citet{BF-HEI} to which the result numbers in the following refer.

\subsection*{Type I solutions}
The type~I solution as given by (\ref{e:I}) has a period of
$2K(m)/\eta$ where $K(m)$ is the complete elliptic integral of the
first kind. From result 340.01 we obtain 
\beq
\alpha_{1/2}=r_1+(r_4-r_1)\frac{\Pi(-\rho|m)}{K(m)}
\eeq
in which $\Pi(n|m)$ is the complete elliptic integral of the third
kind. 

Applying result 340.02 to (\ref{e:I}) yields 
\beqa\nonumber
\alpha_1&=& \frac{(r_4-r_1)^2}{2(\rho+1)(1+m)}\left(
\frac{\rho E(m)+\{\rho^2+2\rho(1+m)+3m\}\Pi(-\rho|m)}{K(m)}
-\rho-m\right) \\
&&\mbox{}+r_1^2+2r_1(r_4-r_1)\frac{\Pi(-\rho|m)}{K(m)}
\eeqa
where $E(m)$ is the complete elliptic integral of the second kind. 
  
\subsection*{Type \IIpe\ solutions}
After introducing
\[
\sigma=\frac{A-B}{A+B}, \qquad
\sigma_1=\frac{Ar\rs{l}-Br\rs{u}}{Ar\rs{l}+Br\rs{u}}, \qquad
\bar{I}_n=
\int_0^{\bar\chi}\!\!\frac{\sod{X}}{(1-\sigma\cn(X|\bar{m}))^n},
\]
the integrals $\alpha_{n/2}$ for $n=1,2,3$ may be written in the form
\[
\alpha_{n/2}=\frac1{\bar\chi}\pfrac{Ar\rs{l}-Br\rs{u}}{A-B}{n}
\sum_{p=0}^n\ba{c}n\\ p\ea (\sigma/\sigma_1-1)^p\bar{I}_p.
\]
From results 341, 
\beq
\bar{I}_1=\frac1{1-\sigma^2}
\left(\Pi(-q;\bar\phi|\bar{m})
+\frac\sigma\mu\tan^{-1}[\mu\sd(\bar\chi|\bar{m})] \right)
\label{e:au1p2pw}
\eeq
where $q=\sigma^2/(1-\sigma^2)$, $\bar\phi=\am(\bar\chi|\bar{m})$, 
$\mu=\sqrt{\bar{m}+q}$, $\sd(x|m)\equiv\sn(x|m)/\dn(x|m)$,
\[
\bar{I}_2=\frac{\{2\bar{m}-(2\bar{m}-1)\sigma^2\}\bar{I}_1
  +\sigma^2E(\bar\phi|\bar{m})-\{\bar{m}+(1-\bar{m})\sigma^2\}\bar\chi
  +\bar\Upsilon_1}{(1-\sigma^2)\{\bar{m}+(1-\bar{m})\sigma^2\}},
\]
\[
\bar{I}_3=\frac{3\{2\bar{m}-(2\bar{m}-1)\sigma^2\}\bar{I}_2
+2\bar{m}F(\bar{\phi}|\bar{m})
-\{6\bar{m}-(2\bar{m}-1)\sigma^2\}\bar{I}_1+\bar\Upsilon_2}
{2(1-\sigma^2)\{\bar{m}+(1-\bar{m})\sigma^2\}},
\]
where $F(\phi|m)$ is the elliptic integral of the first
kind, and
\[
\bar\Upsilon_n=\frac{\sigma^3\sn(\bar\chi|\bar{m})\dn(\bar\chi|\bar{m})}
{(1-\sigma\cn(\bar\chi|\bar{m}))^{n-1}}.
\]
\subsection*{Type \Ipe\ solutions}
The $\alpha_{n/2}$ for $n=1,2,3$ may be written in
the form
\[
     \alpha_{n/2} = \dsfrac{r_1^n}{\chi}\sum_{p=0}^n\ba{c}n\\ p\ea
     \left(r_4/r_1-1 \right)^p I_p,
\] 
where 
\[
I_n=\int_0^{\chi}\!\!\frac{\sod{X}}{(1+\rho\sn^2(X|m))^n}. 
\]      
From result 400.01, $I_1=\Pi(-\rho;\phi|m)$, where $\phi=\am(\chi|m)$.
Results 336.01 and 336.02 yield
\beq
I_2=\frac{\sigma E(\phi|m)-(m+\rho)\chi
+\{2\rho(1+m)+3m+\rho^2\}I_1+\Upsilon_1}{2(1+\rho)(m+\rho)
},
\label{e:I2}
\eeq
\beq
I_3=\frac{mF(\phi|m)-2\{\rho(1+m)+3m\}I_1+3\{2\rho(1+m)+3m+\rho^2\}I_2
  +\Upsilon_2}{4(1+\rho)(m+\rho)},   
\label{e:I3}    
\eeq
respectively, where  
\[ 
\Upsilon_n=
\frac{\rho^2\sn(\chi|m)\cn(\chi|m)\dn(\chi|m)}{(1+\rho\sn^2(\chi|m))^n}. 
\]

\end{document}